\documentclass[twocolumn,showpacs,amsmath,amssymb]{revtex4-1}
\usepackage{bm}
\usepackage[T1]{fontenc}
\usepackage{hyperref}
\input{epsf}

\usepackage{graphicx}
\usepackage{epstopdf}
\usepackage{array}
\usepackage{longtable}
\usepackage{rotating,booktabs}
\usepackage{booktabs,threeparttable}
\usepackage{bm}
\usepackage{float}
\usepackage{amsmath}
\usepackage{gensymb}
\usepackage{multirow}
\usepackage{epsfig}
\usepackage{color}

\begin{document}

\title{Magic wavelengths and triple magic trapping conditions for $5s^2~^1\!S_0$ and $5s5p~^3\!P_{0,2}$ states of Sr atoms}
\author{Yan-Min Wang$^{1}$}
\thanks{These authors contributed equally to this work.}
\author{Qing-Yi Liu$^{1}$}
\thanks{These authors contributed equally to this work.}
\author{Yong-Bo Tang$^{2}$}
\email {tangyongbo@sztu.edu.cn}
\author{Lei Wu$^{1}$}
\author{Deng-Hong Zhang$^{1,3}$}
\author{Chen-Zhong Dong$^{1}$}
\author{Jun Jiang$^{1}$}
\email {phyjiang@yeah.net}

\affiliation{$^{1}$Key Laboratory of Atomic and Molecular
	Physics and Functional Materials of Gansu Province,
	College of Physics and Electronic Engineering,
	Northwest Normal University, Lanzhou 730070, China}

\affiliation{$^{2}$Physics Teaching and Experiment Center, Shenzhen Technology University, Shenzhen 518118, China}
\affiliation{$^{3}$College of Electronic Information and Electronic Engineering, Tianshui Normal University, Tianshui 741000, China}

\date{\today}

\begin{abstract}
The static and dynamic electric dipole polarizabilities of the $5s^2~^1\!S_0$ and $5s5p~^3\!P_{0,2}$ states of Sr atoms are calculated using the relativistic configuration interaction plus the many-body perturbation theory (RCI+MBPT) method. Magic wavelengths are determined for the transitions $5s^2~^1\!S_0\rightarrow 5s5p~^3\!P_{0}$, $5s^2~^1\!S_0\rightarrow 5s5p~^3\!P_{2}$, and $5s5p~^3\!P_0\rightarrow 5s5p~^3\!P_{2}$. A comprehensive study is conducted on the dependence of magic wavelengths on the angle between the laser polarization and the magnetic field. Furthermore, the conditions for realizing triple magic trapping at 813.4~nm for the $5s^2~^1\!S_0$, $5s5p~^3\!P_{0}$ and $5s5p~^3\!P_{2}$ states are investigated. In the case of linearly polarized light, when the angle ($\theta_p$) between the laser polarization direction and the magnetic field is $79.1(0.7)^\circ$, triple magic trapping for the $5s^2~^1\!S_0$, $5s5p~^3\!P_{0}$, and $5s5p~^3\!P_{2}~M=0$ states can be achieved. This result agrees well with the recent experimental measurement (78.49(3)$^\circ$)[Phys. Rev. Lett. 135, 143401 (2025)]. Meanwhile, triple magic trapping involving the $5s5p~^3\!P_{2}~M=2$ state can be achieved when $\theta_p= 37.4(0.3)^\circ$. The conditions for achieving triple magic trapping with circularly and arbitrarily elliptically polarized light are also presented.
\end{abstract}

\maketitle

\section{INTRODUCTION}

The transitions $ns^2~^1\!S_0$ $\rightarrow$ $nsnp~^3\!P_0$ in alkaline earth (-like) atoms (Mg, Ca, Sr, Yb, Hg, Al$^+$, In$^+$, etc.) exhibit narrow linewidths and are insensitive to magnetic fields. These transitions are suitable for high-precision measurements~\cite{ludlow2015}. Based on the transition of $5s^2~^1\!S_0~\rightarrow~5s5p~^3\!P_0$, the Sr optical lattice clock has achieved a record-low systematic uncertainty of 10$^{-19}$ in fractional frequency unit~\cite{aeppli2024,kim2023evaluation,bothwell2022}, making it the most precise neutral atomic clock to date. Moreover, the metastable state $^3\!P_0$ serves as a carrier for qubits, offering numerous opportunities for encoding high-quality qubits and enabling high-fidelity single-qubit gates~\cite{schine2022,Heinz2020, pucher2024,unnikrishnan2024,madjarov2020,lis2023}.

The $5s5p$ $^3\!P_2$ state of Sr atoms is another long-lived metastable state with a lifetime of several minutes~\cite{derevianko2001,yasuda2004, liu2007, Lu2022,klusener2024}. The transition of $5s^2~^1\!S_0\rightarrow 5s5p~^3\!P_2$ can serve as a second clock transition~\cite{klusener2024,onishchenko2019}, with a quality factor of up to $10^{17}$. Interrogating these two distinct clock transitions within a single atomic clock is advantageous for evaluating systematic shifts in clock frequency. Furthermore, the magnetic sublevels of $^3\!P_2$ with $M=\pm1,\pm2$ are sensitive to magnetic fields. Their tunability under external magnetic fields can be used for high-fidelity encoding and local addressing, enabling advanced quantum computing applications~\cite{madjarov2020, lis2023, okuno2022, pucher2024,chen2025}. The combination of the metastable states $^3\!P_0$ and $^3\!P_2$ provides a pathway for implementing high-fidelity two-qubit gates based on fine structure~\cite{pucher2024,unnikrishnan2024}. When combined with the ground state $^1\!S_0$, this system supports the implementation of all-optical qutrits, which offer enhanced computational power and information-carrying capacity.

Recently, coherent control of the fine-structure qubit encoded in the two metastable states, $^3\!P_2$ and $^3\!P_0$, of a single Sr atom has been demonstrated by Pucher \emph{et al.}~\cite{pucher2024} and Unnikrishnan \emph{et al.}~\cite{unnikrishnan2024}. These results pave the way for the development of fast quantum information processors and highly tunable quantum simulators utilizing two-electron atoms. Trautmann \emph{et al.}~\cite{Trautmann2023} studied a qubit encoding scheme that employs the magnetic quadrupole transition between the $^1\!S_0$ and $^3\!P_2$ states of Sr. The angular dependence of this transition was investigated. However, quantitative analysis of the transition strengths in the general case is challenging because each spectral line is broadened and shifted by the differential AC Stark shifts of the $^1\!S_0$ and $^3\!P_2$ states induced by the trapping lattice.

Magic-wavelength trapping is essential for eliminating the AC Stark effect, accurately simulating the angular distribution of transition intensities, and improving coherence time and fidelity. Trautmann \emph{et al.}~\cite{Trautmann2023} also predicted a triple magic condition for both metastable and ground states. Under the triple-magic condition, the $^3\!P_2$, $^3\!P_0$, and $^1\!S_0$ states can experience the same Stark shift. Recently, Maximilian \emph{et al.}~\cite{ammenwerth2025realization} conducted the first experiment demonstrating a triple-magic all-optical qutrit using 813-nm linearly polarized light in $^{88}$Sr atoms. These experiments pave the way for new advancements in qutrit-based quantum metrology.

In this work, the static and dynamic polarizabilities of the $5s^2~^1\!S_0$, $5s5p~^3\!P_0$, and $5s5p~^3\!P_2$ states of Sr atoms are calculated using the relativistic configuration interaction plus many-body perturbation theory (RCI+MBPT) method. The magic wavelengths for the transitions $^1\!S_0~\rightarrow~^3\!P_0$, $^1\!S_0~\rightarrow~^3\!P_2$, and $^3\!P_0~\rightarrow~^3\!P_2$, as well as their variations with respect to the angle between the laser and the magnetic field, are investigated. Subsequently, the conditions for achieving triple magic trapping of the $^1\!S_0$, $^3\!P_0$, and $^3\!P_2$ states at 813.4~nm under linearly, circularly, and arbitrarily elliptically polarized light are presented. Atomic units are used throughout this paper unless otherwise specified.

\begin{figure}[tbh]	
	\centering{
		\includegraphics[width=7cm, height=8.0cm]{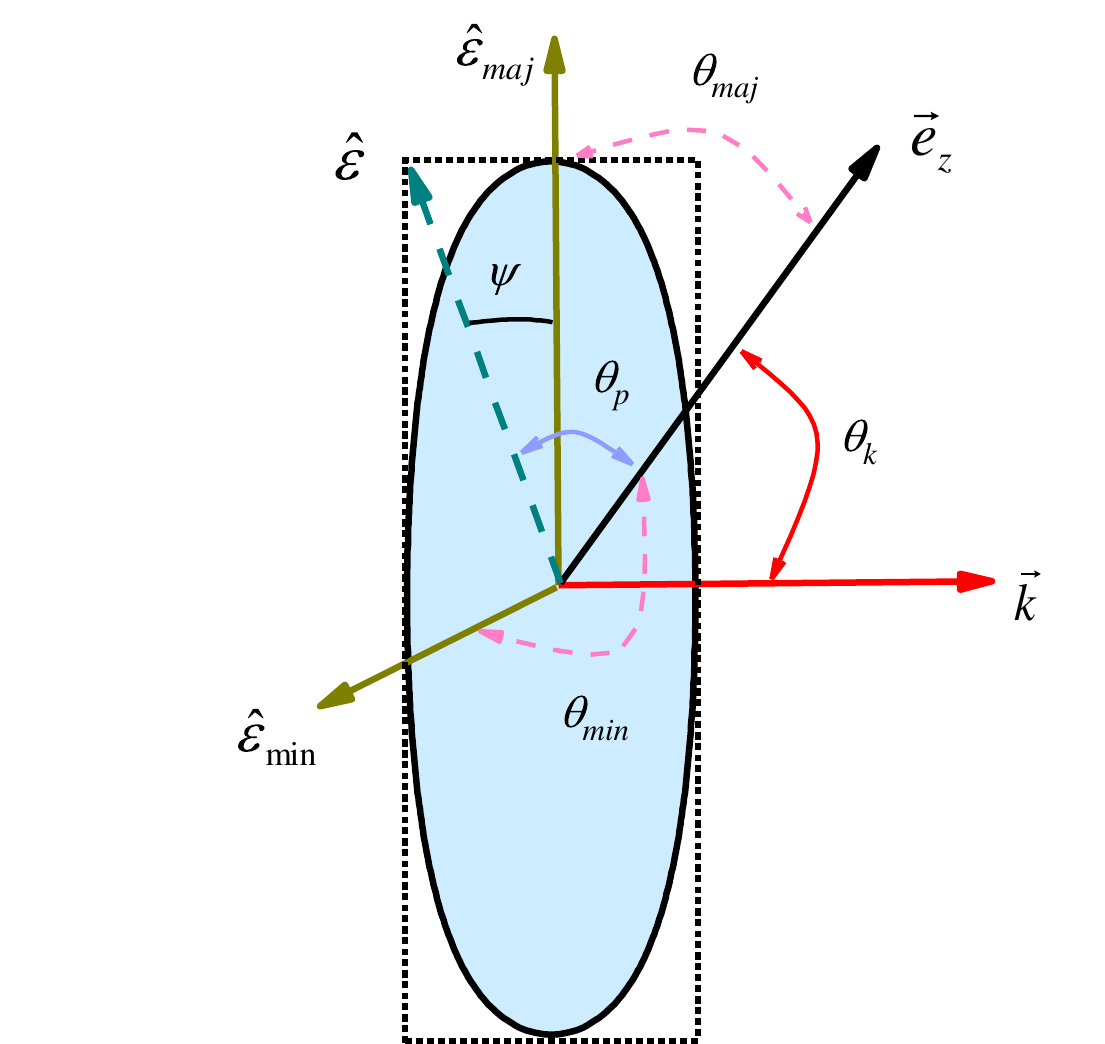}}
\caption{\label{fig1}The schematic of the geometrical parameters of the electromagnetic plane wave and the 1D optical lattice trap. The elliptical area is swept by the electric field vector in one period. The unit vector $\hat{\varepsilon}_\text{maj}$ ($\hat{\varepsilon}_\text{min}$) is aligned with the semimajor (-minor) axis of the ellipse. $\vec{e}_\text{z}$ is the quantization axis, which determines the direction of the magnetic field in the experiment. $\vec{k}$ is the direction of the wave vector. $\theta_p$ is the angle between the polarization vector $\hat{\varepsilon}$ and the quantization axis $\vec{e}_\text{z}$, $\theta_k$ is the angle between $\vec{e}_\text{z}$ and $\vec{k}$. The$\hat{\varepsilon}_\text{maj}$, $\hat{\varepsilon}_\text{min}$, and $\vec{k}$ are orthogonal to each other. $\theta_\text{maj}$ ($\theta_\text{min}$) is the angle between $\hat{\varepsilon}_\text{maj}$ ($\hat{\varepsilon}_\text{min}$) and $\vec{e}_\text{z}$. $\psi$ is directly related to the degree of circular polarization of the electromagnetic plane wave.}
\end{figure}

\section{Theory}\label{Sec.II}

\subsection{RCI+MBPT method}\label{Sec.II-A}

The effective Hamiltonian for the divalent atomic system can be expressed as
\begin{equation}
	\small
    \label{eqadd1}
	\left\{\sum^{2}_{i}{\left[H_{\rm DF}(r_{i})+\Sigma_{1}^{(2)}(r_{i})\right]}+{V_{12}}\right\}|\Psi(\pi JM)\rangle=E|\Psi(\pi JM)\rangle.
\end{equation}
In this expression, $\Sigma_{1}^{(2)}$ is one-body correlation potential constructed via second-order perturbation theory~\cite{dzuba2006calculations}, and $H_{\rm DF}$ denotes the single-electron DF Hamiltonian, expressed as:
\begin{equation}
	\label{eqadd2}
	H_{DF}=c\alpha_i\cdot\mathbf{p}_i+(\beta_i-1)c^2+V_N(r_i)+V_{\mathrm{DF}},
\end{equation}
\ where $\alpha_i$ and $\beta_i$ denote Dirac matrices, $\mathbf{p}_i$ represents the momentum operator of an electron, $V_N(r_i)$ is the Coulomb potential between the valence electron and the atomic nucleus, $V_{\mathrm{DF}}$ is the DF potential. The two-body interaction Hamiltonian is expressed as:
\begin{equation}
	\label{eqadd3}
	V_{12}=\frac{1}{r_{12}}+\Sigma_{2}^{(2)},
\end{equation}
which includes the Coulomb interaction potential $\frac{1}{r_{12}}$ between electrons and the two-body correlation potential $\Sigma_{2}^{(2)}$ constructed by second-order perturbation theory.

The system wavefunction $|\Psi(\pi JM)\rangle$ can be expressed as a linear combination of configuration wavefunctions $|\Psi_{kl}(\pi JM)\rangle$ with the same parity $\pi$, angular momentum $J$, and magnetic quantum number $M$, namely:
\begin{equation}
	\label{eqadd4}
	|\Psi(\pi JM)\rangle=\sum_{k\leqslant l}C_{kl}|\Phi_{kl}(\pi JM)\rangle,
\end{equation}
where $C_{kl}$ denotes the expansion coefficient, the subscripts of $k$ and $l$ represent valence electron orbitals. The configuration wavefunction $|\Phi_{kl}(\pi JM)\rangle$ is composed of a single determinant basis constructed from single-particle orbitals and can be expressed as:
\begin{equation}
	\label{eqadd5}
	|\Phi_{kl}(\pi JM)\rangle=\eta_{kl}\sum_{m_{k},m_{l}}\langle j_{k}m_{l},j_{k}m_{l}|JM\rangle a_{k}^{\dagger}a_{l}^{\dagger}|0\rangle.
\end{equation}
where the coefficient $\eta_{kl}$ is defined as:
\begin{equation}
	\label{eqadd6}
	\eta_{kl}=
    \begin{cases}
    \frac{\sqrt{2}}{2}, & \quad k=l, \\
    1, & \quad k\neq l.
    \end{cases}
\end{equation}
Substituting Eqs.~(\ref{eqadd4}) into~(\ref{eqadd1}) and applying the variational principle yields a generalized eigenvalue equation.
\begin{equation}
	\label{eqadd7}
    \sum_{x < y} [(H_{DF}+\Sigma_{1}^{(2)})_{kl,xy} + (V_{12})_{kl,xy}] C_{xy} = EC_{kl},
\end{equation}
where the subscripts of $x$ and $y$ also represent valence electron orbitals. By diagonalizing this equation, the wave function and energy of the target state can be obtained. The matrix elements for the one-body Hamiltonian are given by:
\begin{equation}
	\label{eqadd8}
	\begin{aligned}
		&\langle\Phi_{xy}(JM)|H_{DF}+\Sigma_{1}^{(2)}|\Phi_{kl}(JM)\rangle=(\varepsilon_{x}+\varepsilon_{y})\delta_{xk}\delta_{yl} \\
		&+\eta_{xy}\eta_{kl}\left\{(\Sigma_{1}^{(2)})_{xk}\delta_{yl}+(\Sigma_{1}^{(2)})_{yl}\delta_{xk}\right. \\
		&+(-1)^{J}\left.\left[(\Sigma_{1}^{(2)})_{xl}\delta_{yk}+(\Sigma_{1}^{(2)})_{yk}\delta_{xl}\right]\right\}.
	\end{aligned}
\end{equation}
Here, $(\Sigma_{1}^{(2)})_{kl}$ can be expressed as~\cite{Safronova2009pra}:
\begin{equation}
\label{eqadd9}
\begin{aligned}
(\Sigma_{1}^{(2)})_{kl} & =-\sum_{rab}\sum_{L}\frac{\delta_{\kappa_{k}\kappa_{l}}}{[L][j_{k}]}\frac{X_{L}(rkab)Z_{L}(rlab)}{\varepsilon_{ab}-\varepsilon_{kr}+\tilde{\varepsilon}_{l}-\varepsilon_{l}} \\
 & +\sum_{rsa}\sum_{L}\frac{\delta_{\kappa_{k}\kappa_{l}}}{[L][j_{k}]}\frac{X_{L}(rsla)Z_{L}(rska)}{\tilde{\varepsilon}_{k}+\varepsilon_{a}-\varepsilon_{rs}},
\end{aligned}
\end{equation}
where $\varepsilon_{a}$ is the single-particle energy, $\varepsilon_{ab}$ = $\varepsilon_{a}$ + $\varepsilon_{b}$, and $\tilde{\varepsilon}_{k}$ is the Dirac-Fock energy of the lowest valence state of a given symmetry. $a$ and $b$ denote atomic core orbitals, while $r$ and $s$ are designated as the virtual states (or called excited orbitals). $Z_{L}$ and $X_{L}$ can be expressed as~\cite{Safronova2009pra}:
\begin{equation}
	\small\label{eqadd11}
Z_{L}(rsab)=X_{L}(rsab)+\sum_{K'}[L]
\begin{Bmatrix}
j_{r} & j_{a} & L \\
j_{s} & j_{b} & K'
\end{Bmatrix}X_{K'}(rsba),
\end{equation}
and
\begin{equation}
	\small\label{eqadd12}
X_{L}(rsab)=(-1)^{L}\langle\kappa_{r}\parallel C^{L}\parallel\kappa_{a}\rangle\langle\kappa_{s}\parallel C^{L}\parallel\kappa_{b}\rangle R_{L}(rsab),
\end{equation}
$\langle\kappa_{r}\parallel C^{L}\parallel\kappa_{a}\rangle$ represents the reduced matrix element, and $R_{L}(rsab)$ denotes the Slater integral, which can be expressed as~\cite{Safronova2009pra}:
\begin{equation}
	\label{eqadd13}
\begin{aligned}
\langle k_{r}\|C^{L}\|k_{a}\rangle & =(-1)^{j_{r}+\frac{1}{2}}\sqrt{(2j_{r}+1)(2j_{a}+1)} \\
 & \times
\begin{pmatrix}
\mathrm{j_r} & \mathrm{j_a} & \mathrm{k} \\
-\frac{1}{2} & \frac{1}{2} & 0
\end{pmatrix}\prod(\ell_\mathrm{r}+\ell_a+\mathrm{k}),
\end{aligned}
\end{equation}
and
\begin{equation}
	\small\label{eqadd14}
\begin{aligned}
R_{L}(rsab) & =\int dr_{1}[P_{r}(r_{1})P_{a}(r_{1})+Q_{r}(r_{1})Q_{a}(r_{1})] \\
 & \times\int dr_{2}\frac{r_{<}^{k}}{r_{>}^{k+1}}[P_{s}(r_{2})P_{b}(r_{2})+Q_{s}(r_{2})Q_{b}(r_{2}),
\end{aligned}
\end{equation}
where $P$ and $Q$ represent the large and small components of the radial wave function, respectively. The matrix form of the two-body interaction Hamiltonian $V_{12}$ can be expressed as:
\begin{equation}
	\begin{aligned}
		&\langle\Phi_{xy}(JM)|V_{12}|\Phi_{kl}(JM)\rangle \\
		&= \eta_{xy}\eta_{kl}\sum\left\{(-1)^{(J+L+j_y+j_k)}\begin{Bmatrix}j_x & j_y & J\\j_l & j_k & L\end{Bmatrix}\right. \\
		&\quad\times\left.\left[X_L(xykl)+(\Sigma_2^{(2)})_L(xykl)\right]\right. \\
		&\quad+(-1)^{(L+j_y+j_k)}\begin{Bmatrix}j_x & j_y & J\\j_k & j_l & L\end{Bmatrix} \\
		&\quad\times\left.\left[X_L(xylk)+(\Sigma_2^{(2)})_L(xylk)\right]\right\},
	\end{aligned}
\end{equation}
where $(\Sigma_{2}^{(2)})_{L}(xykl)$ can be expressed as~\cite{Safronova2009pra}:
\begin{equation}
	\label{eqadd10}
	\begin{aligned}
		(\Sigma_{2}^{(2)}) & _{L}(xykl)= \\
		& \left.\sum_{ab}\sum_{KK^{\prime}}\left[L\right]\left\{
		\begin{array}
			{ccc}j_{x} & j_{k} & L \\
			K & K^{\prime} & j_{a}
		\end{array}\right.\right\}
		\begin{Bmatrix}
			j_{y} & j_{l} & L \\
			K & K^{\prime} & j_{b}
		\end{Bmatrix} \\
	   &\times\frac{X_{K}(klab)X_{K^{\prime}}(abkl)}{\varepsilon_{ab}-\varepsilon_{xy}+\tilde{\varepsilon}_{k}+\tilde{\varepsilon}_{l}-\varepsilon_{k}-\varepsilon_{l}} \\
		& -\sum_{ra}\frac{(-1)^{j_{k}+j_{x}+L}}{[L]}\frac{Z_{L}(lrya)Z_{L}(xrka)}{\varepsilon_{a}-\varepsilon_{mr}+\tilde{\varepsilon}_{k}+\tilde{\varepsilon}_{l}-\varepsilon_{l}} \\
		& -\sum_{ra}\frac{(-1)^{j_{k}+j_{x}+L}}{[L]}\frac{Z_{L}(krxa)Z_{L}(yrka)}{\varepsilon_{a}-\varepsilon_{nr}+\tilde{\varepsilon}_{k}+\tilde{\varepsilon}_{l}-\varepsilon_{k}},
	\end{aligned}
\end{equation}
$[L]= 2L + 1$, where $L$ denotes the order.

%
%

When calculating the one-body potential part, we incorporate a scaling factor $\rho_{\kappa}$, which is applied as $\rho_{\kappa}\Sigma_{1}$. This factor serves a dual purpose. On the one hand, it compensates for higher-order effects beyond the second-order MBPT. On the other hand, it expedites the convergence of energy and matrix elements within the RCI+MBPT method. As a result, substantial computational resources can be conserved. The scaling factor is determined by fitting the band energy of the single-electron system, using a method similar to the DFCP approach~\cite{tang2013pra}.

When calculating the transition matrix element, it is essential to consider the core polarization correction. In the present work, we adopted the Random Phase Approximation (RPA) to incorporate this correction, as referenced in the works of Dzuba \textit{et al.} and Savukov\textit{ et al.}~\cite{Dzuba1998JETP,Savukov2000pra}. Further details on the RCI+MBPT method can be found in Refs.~\cite{Dzuba1996pra,Dzuba1998pra,Porsev2001pra,Savukov2002pra,Safronova2009pra,zhang2023}.

In the DF calculation, the large and small components of the Dirac wave functions are expanded using $50$ B-spline bases of order $k=17$, and the box size $R_\text{max}=120$~$a_0$. The partial waves are limited to $\ell_\text{max} = 6$ and the lowest $40$ orbitals for each partial wave are used to construct the configuration space. In the second-order many-body perturbation calculations, the summation is performed over the entire basis set.

\subsection{Polarizabilities}

\begin{table*}
	\caption []{Comparison of the energy levels (in cm$^{-1}$) for some low-lying states: "CI+MBPT" and "CI+All" represent the result obtained by using the configuration interaction plus many-body perturbation theory and the configuration interaction plus all-order linearized coupled-cluster method in Ref.~\cite{Safronova2013}. All energies are given relative to the ground state of the Sr$^{2+}$ core. The relative differences between the present energy and the NIST energy~\cite{NIST_ASD} are listed as Diff.}\label{Tab2}
	\begin{ruledtabular}
		\begin{tabular}{cccccc}
			{State} & {Present} & CI+MBPT~\cite{Safronova2013}  & CI+All~\cite{Safronova2013} & {NIST~\cite{NIST_ASD}} &  Diff.  \\			
			\midrule
			$5s^2~^1\!S_{0}$ & $-$135027.1 & $-$136244 & $-$135444 & $-$134897.4 & 0.10\% \\
			$5s5p~^3\!P^o_{0}$ & $-$120679.6 & $-$121438 & $-$120894 & $-$120579.9 & 0.08\% \\
			$5s5p~^3\!P^o_{1}$ & $-$120496.3 & $-$121249 & $-$120705 & $-$120393.0 & 0.09\%\\
			$5s5p~^3\!P^o_{2}$ & $-$120104.2 & $-$120845 & $-$120302 & $-$119998.8 & 0.09\%\\
            $5s4d~^3\!D_{1}$ & $-$117011.2 & $-$118019 & $-$117117 & $-$116738.3 & 0.23\%\\
            $5s4d~^3\!D_{2}$ & $-$116949.2 & $-$117946 & $-$117050 & $-$116678.6 & 0.23\%\\
            $5s4d~^3\!D_{3}$ & $-$116846.1 & $-$117822 & $-$116938 & $-$116578.1 & 0.23\%\\
            $5s4d~^1\!D_{2}$ & $-$114866.8 & $-$115816 & $-$115003 & $-$114747.7 & 0.10\%\\
			$5s5p~^1\!P^o_{1}$ & $-$113530.9 & $-$114289 & $-$113621 & $-$113198.9 & 0.29\%\\
			$5s6s~^3\!S_{1}$ & $-$106120.2 & $-$106875 & $-$106221 & $-$105858.6 & 0.25\%\\
			$5s6s~^1\!S_{0}$ & $-$104552.3 & $-$105306 & $-$104667 & $-$104305.5 & 0.24\%\\
            $4d5p~^3\!F^o_{2}$ & $-$101943.8 & $-$102525 & $-$101796 & $-$101630.5 & 0.31\%\\
            $4d5p~^3\!F^o_{3}$ & $-$101611.4 & $-$102155 & $-$101441 & $-$101307.7 & 0.30\%\\
            $4d5p~^1\!D^o_{2}$ & $-$101424.6 & $-$102026 & $-$101236 & $-$101070.5 & 0.35\%\\
			$5s6p~^3\!P^o_{0}$ & $-$101303.9 & $-$102003 & $-$101389 & $-$101043.9 & 0.26\%\\
			$5s6p~^3\!P^o_{1}$ & $-$101291.5 & $-$101989 & $-$101373 & $-$101029.1 & 0.26\%\\
			$5s6p~^3\!P^o_{2}$ & $-$101189.4 & $-$101879 & $-$101310 & $-$100924.3 & 0.26\%\\
			$5s6p~^1\!P^o_{1}$ & $-$101081.0 & $-$101768 &  $-$101136 & $-$100799.0 & 0.28\% \\
            $5s5d~^1\!D_{2}$ & $-$100449.5 & $-$101152 & $-$100486 & $-$100169.9 & 0.28\%\\
			$5s5d~^3\!D_{1}$ & $-$100159.2 & $-$100873 & $-$100234 & $-$99890.5 & 0.27\%\\
			$5s5d~^3\!D_{2}$ & $-$100144.1 & $-$100856 & $-$100218 & $-$99875.4 & 0.27\%\\
                $5p^2~^3\!P_{0}$ & $-$99937.1 & $-$100390 & $-$99899 & $-$99703.9 & 0.23\%\\
                $5p^2~^3\!P_{1}$ & $-$99727.5 & $-$100174 & $-$99686 & $-$99497.3 & 0.23\%\\
                $5p^2~^3\!P_{2}$ & $-$99453.7 & $-$99900 & $-$99405 & $-$99222.7 & 0.23\%\\
                $4d5p~^3\!D^o_{1}$ & $-$99066.7&&& $-$98633.2  & 0.44\%  \\
                $4d5p~^3\!D^o_{2}$ & $-$98947.5&&& $-$98515.6 & 0.44\%  \\
                $4d5p~^3\!D^o_{3}$ & $-$98764.6&&& $-$98337.9 & 0.43\%  \\
                $4d^2~^1\!D_{2}$ & $-$98239.5&&& $-$97936.5  & 0.31\%  \\
                $5p^2~^1\!S_{0}$ & $-$98022.1&&& $-$97737.1  & 0.29\%  \\
                $4d5p~^3\!P^o_{0}$ & $-$97967.9&&& $-$97605.3  & 0.37\%  \\
                $4d5p~^3\!P^o_{1}$ & $-$97953.6&&& $-$97594.6 & 0.37\%  \\
                $4d5p~^3\!P^o_{2}$ & $-$97913.4&&& $-$97560.8 & 0.36\%  \\
                $5s7s~^3\!S_{1}$ & $-$97761.2 & $-$98468 & $-$97838 & $-$97472.7 & 0.30\%\\
                $4d5p~^1\!F^o_{3}$ & $-$97217.5&&& $-$96889.6& 0.34\%  \\
                $5s7s~^1\!S_{0}$ & $-$96692.8&&& $-$96453.4  & 0.25\% \\
			$5s7p~^3\!P^o_{0}$ & $-$95787.0&&& $-$95485.7 & 0.32\% \\
                $5s4f~^1\!F^o_{3}$ & $-$95702.3&&& $-$95358.4 & 0.36\% \\
                $5s6d~^3\!D_{1}$ & $-$95500.9 & $-$96194 & $-$95568 & $-$95211.5 & 0.30\%\\
                $5s8s~^1\!S_{0}$ & $-$94128.1&&& $-$93845.0  & 0.30\%  \\
                $5s8p~^3\!P^o_{0}$ & $-$93483.8&&& $-$93185.3 & 0.32\% \\
                $5s9p~^3\!P^o_{0}$ & $-$92211.3&&& $-$91911.5 & 0.33\% \\
		\end{tabular}
	\end{ruledtabular}				
\end{table*}

\begin{table*}
\caption[]{Comparison of some reduced $E1$ matrix elements (in a.u.) of Sr. The numbers in the parentheses are uncertainties.}\label{Tab3}
\begin{ruledtabular}
	\begin{tabular}{clllll}			
		{Transition} & {Present}& {CI+All}& {MBPT}& {Expt.} &
            {NIST~\cite{NIST_ASD}} \\
             \midrule
	$5s^2~^1\!S_{0}\rightarrow5s5p~^3\!P_{1}$ & 0.156(10) &  0.158~\cite{Safronova2013}&
         & \textbf{0.151(2)}~\cite{drozdowski1997} & 0.151\\
         &&0.151~\cite{Porsev2014} \\
	$5s^2~^1\!S_{0}\rightarrow5s5p~^1\!P_{1}$ & 5.303(74) & 5.272~\cite{Safronova2013} &
        5.253~\cite{Safronova2013}   & 5.248(12)~\cite{Heinz2020}& 5.394 \\
	&&& 5.28~\cite{Porsev2008}   & \textbf{5.248(2)}~\cite{yasuda2006}\\
	$5s^2~^1\!S_{0}\rightarrow5s6p~^1\!P_{1}$ & 0.253(23) & 0.281~\cite{Safronova2013} &
        0.236~\cite{Porsev2008}  &\textbf{0.26(2)}~\cite{Parkinson1976} & 0.266\\

        $5s5p~^3\!P_{0}\rightarrow5s4d~^3\!D_{1}$ & 2.708(45) & 2.712~\cite{Safronova2013} & 2.681~\cite{Safronova2013}   &\textbf{2.6707(62)}~\cite{nicholson2015}\\
        & & 2.675(13)~\cite{Safronova2013}& 2.74~\cite{Porsev2008} &  2.7(1)~\cite{Redondo2004}\\
        &&&&2.5(1)~\cite{miller1992} \\
	$5s5p~^3\!P_{0}\rightarrow5s6s~^3\!S_{1}$ & 1.979(33) & 1.970~\cite{Safronova2013}&
        1.983~\cite{Safronova2013}  &  \textbf{1.962(10)}\footnotemark[1]~\cite{Safronova2013} &2.02\\
	&   &   & 1.96~\cite{Porsev2008} &2.03(6)~\cite{1984Jonsson}\\
        $5s5p~^3\!P_{0}\rightarrow5s5d~^3\!D_{1}$ & 2.501(41) & 2.460~\cite{Safronova2013} &  2.474~\cite{Safronova2013}   & \textbf{2.450(24)}\footnotemark[1]~\cite{Safronova2013} &2.35\\
        & &  & 2.50~\cite{Porsev2008} &2.3(1)~\cite{Sansonetti2010} \\
	$5s5p~^3\!P_{0}\rightarrow5p^2~^3\!P_{1}$ & 2.600(43) & 2.619~\cite{Safronova2013} &
         2.587~\cite{Safronova2013}  &\textbf{2.605(26)}\footnotemark[1]~\cite{Safronova2013} &2.49\\
	&   &  &2.56~\cite{Porsev2008} & 2.5(1)~\cite{Sansonetti2010} \\	
		
        $5s5p~^3\!P_{1}\rightarrow5s4d~^3\!D_{2}$ & 4.068(111) & 4.019(20)~\cite{Porsev2014}  & 4.031~\cite{Porsev2014} &\\
        &&4.013(9)~\cite{Kestler2022} \\
        $5s5p~^3\!P_{1}\rightarrow5s5d~^3\!D_{2}$ & 3.737(102)  & 3.673(37)~\cite{Porsev2014} & 3.720~\cite{Porsev2014}   & &3.74\\	
        & &3.671(36)~\cite{Kestler2022}  \\
	$5s5p~^3\!P_{1}\rightarrow5s4d~^3\!D_{1}$ & 2.349(39) & 2.322(11)~\cite{Safronova2013}  &
        2.326~\cite{Porsev2014}  \\
	& &2.318(5)~\cite{Kestler2022}  \\
        $5s5p~^3\!P_{1}\rightarrow5s6s~^3\!S_{1}$ & 3.455(94) &3.425(17)~\cite{Porsev2014}& 3.463~\cite{Porsev2014}   & &3.60\\
        & & 3.435(14)~\cite{Kestler2022}  \\
        $5s5p~^3\!P_{1}\rightarrow5p^2~^3\!P_{0}$ & 2.669(44) & 2.657(27)~\cite{Porsev2014}& 2.619~\cite{Porsev2014} & \\
        &&2.658(27)~\cite{Kestler2022} \\

        $5s5p~^3\!P_{2}\rightarrow5s4d~^3\!D_{3}$ &5.601(77) & 5.614~\cite{Safronova2013} \\
        &&5.530~\cite{Trautmann2023} \\
        $5s5p~^3\!P_{2}\rightarrow5s5d~^3\!D_{3}$ &5.071(70) &4.994~\cite{Trautmann2023}&&& 5.089\\
	$5s5p~^3\!P_{2}\rightarrow5s4d~^3\!D_{2}$ &2.362(39) &2.367~\cite{Safronova2013} \\
        &&2.331~\cite{Trautmann2023} \\
        $5s5p~^3\!P_{2}\rightarrow5s5d~^3\!D_{2}$ & 1.975(33) & 1.956~\cite{Trautmann2023} &&&1.970\\
        $5s5p~^3\!P_{2}\rightarrow5s4d~^1\!D_{2}$ & 0.091(8) & 0.102~\cite{Trautmann2023} & 0.10(2)~\cite{Porsev2001}  	 \\
	$5s5p~^3\!P_{2}\rightarrow5s4d~^3\!D_{1}$ & 0.609(40) &0.612~\cite{Safronova2013} &
        0.604~\cite{Safronova2013} \\
        &&0.6021~\cite{Trautmann2023} \\
        &&0.670~\cite{Safronova2013} \\
        $5s5p~^3\!P_{2}\rightarrow5s6s~^3\!S_{1}$ & 4.546(124) &4.521~\cite{Trautmann2023}&&& 4.69\\
	\end{tabular}
    \footnotetext[1]{These values are derived from a combined theoretical and experimental analysis of Stark shifts and magic wavelengths~\cite{Safronova2013}.}
\end{ruledtabular}	
\end{table*}

When an atom is placed in an external laser field with frequency $\omega$, the Stark effect occurs. The Stark energy shift is related to the atom's polarizability. The polarizability of an atom in state $i$ can be expressed as~\cite{manakov1986,beloy2009,le2013,mitroy2010}:
\begin{eqnarray}\label{eq2}
\alpha_i(\omega)=&&\alpha^S_i(\omega)+\mathcal{A}\text{cos}\theta_k \frac{M_i}{2J_i} \alpha^V_i(\omega)
\nonumber \\
&&+\frac{3\text{cos}^2\theta_p-1}{2}\frac{3M^2_i-J_i(J_i+1)}{J_i(2J_i-1)}\alpha^T_i(\omega),
\end{eqnarray}
where, $\alpha^S_i(\omega)$, $\alpha^V_i(\omega)$, and $\alpha^T_i(\omega)$ represent the scalar, vector, and tensor polarizabilities, respectively. $M_i$ is the component of the total angular momentum $J_i$. $\mathcal{A}$ denotes the degree of polarization ($\mathcal{A}=0$ for linearly polarized light, $\mathcal{A}=+1$ and $\mathcal{A}=-1$ are for right-handed and left-handed circularly polarized light, respectively). As shown in Figure~\ref{fig1}, $\theta_k$ is the angle between the quantization axis (the direction of the static magnetic field) $\vec{e}_\text{z}$ and the wave vector $\vec{k}$, with $\text{cos}\theta_k=\vec{k}\cdot\vec{e}_\text{z}$.
Geometrically, $\theta_p$ is related to the $\theta_{\text{maj}}$ and $\theta_{\text{min}}$ and can be expressed as:
\begin{eqnarray}\label{eq3}
	\text{cos}^2\theta_p&&=\text{cos}^2\psi \text{cos}^2\theta_\text{maj}+\text{sin}^2\psi \text{cos}^2\theta_\text{min}
	\nonumber\\
	&&=\text{sin}^2\psi \text{sin}^2\theta_{k}+\text{cos}(2\psi)\text{cos}^2\theta_\text{maj},
\end{eqnarray}
where $\theta_{\text{maj}}$ ($\theta_{\text{min}}$) is the angle between the major (minor) axis of the ellipse and the $\vec{e}_\text{z}$ axis. For linearly polarized light, $\theta_p$ is the angle between the laser polarization vector $\hat{\varepsilon}$ and the $\vec{e}_\text{z}$ axis. The degree of ellipticity $\mathcal{A}$ is directly related to the angle $\psi$$(\left|\psi\right|\leq 45^\circ)$,
\begin{eqnarray}\label{eq4}
	\mathcal{A}=\text{sin}2\psi.
\end{eqnarray}
$\alpha^S_i(\omega)$, $\alpha^V_i(\omega)$, and $\alpha^T_i(\omega)$ can be calculated by sum-over-states method, that is
\begin{eqnarray}\label{eq5}
\alpha^S_i(\omega)=\frac{2}{3(2J_i+1)}
\sum_{j}\frac{\varepsilon_{ij}{\langle \gamma_j J_j\lVert d\rVert \gamma_i J_i\rangle}^2}{\varepsilon^2_{ij}-\omega^2},
\end{eqnarray}
\begin{eqnarray}\label{eq6}
&&\alpha^V_i(\omega)=-2\sqrt{\frac{6J_i}{(J_i+1)(2J_i+1)}}
\nonumber \\
&&\times\sum_{j}{(-1)^{J_i+J_j}
	\left\{
	\begin{array}{ccc}
		1 & 1 & 1 \\
		J_i & J_i & J_j\\
	\end{array}
	\right\}
	\frac{\omega\langle \gamma_j J_j \lVert d\rVert \gamma_i J_i\rangle^2}{\varepsilon^2_{ij}-\omega^2}},
\end{eqnarray}
and
\begin{align}\label{eq7}
&\alpha^T_i(\omega)= 4\left(\frac{5J_i(2J_i-1)}{6(J_i+1)(2J_i+1)(2J_i+3)}\right)^{1/2}
\nonumber \\
& \times\sum_{j}(-1)^{J_i+J_j}
\left\{
\begin{array}{ccc}
	J_i & 1 & J_j \\
	1 & J_i & 2   \\
\end{array}
\right\}
\frac{{\langle \gamma_j J_j\lVert d\rVert \gamma_i J_i\rangle}^2\varepsilon^2_{ij}}{\varepsilon^2_{ij}-\omega^2},
\end{align}
where $\varepsilon_{ij} = E_j-E_i$is the excitation energy, $\langle \gamma_j J_j \lVert d\rVert \gamma_i J_i\rangle$ is the reduced $E1$ transition matrix element. If $\omega=0$, the dynamic polarizabilities in Eqs.~(\ref{eq5}) and (\ref{eq7}) reduce to static polarizabilities, and the static vector polarizability equals zero.

The calculation of the dynamic polarizability of the Sr$^{2+}$ atomic core adopts the pseudo-oscillator strength distribution method~\cite{Jiang2016}.

\section{results and discussion} \label{Sec.III}

\subsection{Energy levels, reduced $E1$ matrix elements and static polarizabilities}

Table~\ref{Tab2} presents some of the low-lying energy levels of the present calculations with results from other theoretical methods, including the configuration interaction plus many-body perturbation theory (CI+MBPT)~\cite{Safronova2013}, the configuration interaction plus all-order linearized coupled-cluster (CI+All)~\cite{Safronova2013}, and the NIST Atomic Spectra Database~\cite{NIST_ASD}. Our results show good agreement with the NIST data. In particular, the differences of $5s^2~^1\!S_0$ and $5s5p~^3\!P_{0,1,2}$ are within 0.10\%. The maximum difference for other excited states does not exceed 0.5\%. Overall, the results represent an improvement over those reported in~\cite{Safronova2013}. Our results are consistent with those obtained using the Cl+All method in~\cite{Safronova2013}, with differences generally within 0.2\%.

 \begin{table*}
 	\caption[]	{Comparison of the static dipole polarizabilities (a.u.) of the low-lying states of Sr with available experimental and theoretical results.  The polarizability of the core is $\alpha_{\text{core}}=5.29 $ a.u.~\cite{Safronova2013}.}\label{Tab4}
 	\begin{ruledtabular}
 		\begin{tabular}{cllllll}
 			& \multicolumn{3}{c}{$\alpha^S$} & \multicolumn{3}{c}{$\alpha^T$} \\
                \cmidrule(r){2-4} \cmidrule(r){5-7}
 			 \multicolumn{1}{c}{State} & \multicolumn{1}{c}{Present} & \multicolumn{1}{c}{Theory} & \multicolumn{1}{c}{Expt.} & \multicolumn{1}{c}{Present} & \multicolumn{1}{c}{Theory} & \multicolumn{1}{c}{Expt.}\\
 			\midrule
$5s^2~^1\!S_0$ & 201.2(3.3) &  197.14(20) RCI+All~\cite{Safronova2013}
                        &186(15)~\cite{molof1974}\\
                        & \textbf{197.6(1.0)}\footnotemark[2] & 197.2 CI+MBPT~\cite{Porsev2008} \\
                        && 197.2(2) Hybrid-Sum rule~\cite{porsev2006high}\\
                        && 197.8 CICP~\cite{Cheng2013}\\
                        && 198.85 RCCSD~\cite{lim1999}\\
 		            && 199.7(1.2)RCC~\cite{yu2015}\\
                        && 214.5 MCDF~\cite{shukla2020}\\
$5s5p~^3\!P_0$ & 454.1(4.5) & 444.51(20) CI+All~\cite{Safronova2013}&    \\
                      & \textbf{443.9(3.3)}\footnotemark[2] & 444.1(2.7) RCC~\cite{yu2015}\\
                      && 457.0 CI+MBPT~\cite{Porsev2008}\\
                      && 458.1 CI+All~\cite{Porsev2014}\\

$5s5p~^3\!P_1$ & 469.0(1.7) & 459.2(3.8)CI+All ~\cite{Porsev2014}  & &  25.2(0.7) & 26.0(2)
 CI+All~\cite{Porsev2014}& 24.5~\cite{oppen1969}\\
                      &  &497.0 CICP~\cite{mitroy2010dispersion} &&&
                      27.7CICP~\cite{mitroy2010dispersion}  \\
$5s5p~^3\!P_2$ & 504.3(2.9) & 491.1(1.0)  RCC~\cite{yu2015} &&{$-54.9(1)$} &{$-$54.5(2.8)} RCC~\cite{yu2015}\\
             &&&&  &$-$48.56~\cite{Trautmann2023}\\
\end{tabular}
\end{ruledtabular}
\footnotetext[2]{In calculating these values, some of the matrix elements we calculated were replaced with experimental results, i.e., the bold data in Table~\ref{Tab3}.}

\end{table*}

Table~\ref{Tab3} lists several dipole transition matrix elements for the $5s^2~^1\!S_{0}$ and $5s5p~^3\!P_{0,1,2}$ states. These results are compared with existing experimental and theoretical results. For transitions from the ground state $5s^2~^1\!S_0$, our calculated results are approximately 3\% larger than the experimental results~\cite{drozdowski1997, Heinz2020, yasuda2006}. For the states $5s5p~^3\!P_{0,1}$, our results show good agreement with those calculated using the CI+All method~\cite{Safronova2013,Porsev2014,Kestler2022} and the many-body perturbation theory (MBPT) in Refs.~\cite{Safronova2013, Porsev2008}, with a difference of 1.5\%. For the $5s5p~^3\!P_{2}$ state, the discrepancy between our results and those in Ref.~\cite{Safronova2013} is less than 1\%. Additionally, all current values are consistent with the NIST tabulations~\cite{NIST_ASD}, which were derived from the line strengths.

\begin{figure}[tbh]	
	\centering{
		\includegraphics[width=8cm, height=10.0cm]{1S0-3P0.pdf}}
	\caption{\label{fig2}Dynamical polarizabilities of the $5s^2~^1\!S_{0}$ and $5s5p~^3\!P_0$ states. The dashed vertical lines indicate the positions of the resonant transitions. The magic wavelengths are identified by arrows where the polarizabilities of $5s5p~^3\!P_0$ and $5s^2~^1\!S_{0}$ are equal.}
\end{figure}

\begin{table*}[]
	\caption[]{Contributions to the $5s^2~^1\!S_0$ and $5s5p~^3\!P_0$ static polarizabilities of Sr (in a.u.).}
	\label{tab:contribution}
    \begin{ruledtabular}
		\begin{tabular}{clllll}
			Initial State & Final State & \multicolumn{2}{c}{Method\footnotemark[3]} & \multicolumn{2}{c}{Method\footnotemark[4]} \\
            \cmidrule(r){3-4} \cmidrule(r){5-6}
            &&\multicolumn{1}{c}{Theoretical matrix elements} &\multicolumn{1}{c}{Contribution}&\multicolumn{1}{c}{Theoretical matrix elements}&\multicolumn{1}{c}{Contribution}\\
            \midrule
			$5s^2~^1\!S_0$ & $5s5p~^3\!P_1$ & 0.156(10) & 0.248 & 0.151(2)~\cite{drozdowski1997} & 0.23\\
            & $5s5p~^1\!P_1$ & 5.303(73) & 189.67 & 5.248(2)~\cite{yasuda2006} & 185.86\\
			  & $5s6p~^1\!P_1$ & 0.253(23) & 0.275 & 0.26(2)~\cite{Parkinson1976} & 0.29\\
            & Other &  & 4.68 &  & 4.68\\
            &$\alpha_{\text{core}}$ &  & 5.29 &  & 5.29\\
            & Total &  & 201.2 &  & 197.6\\
        \midrule
        $5s5p~^3\!P_0$ & $5s4d~^3\!D_1$ & 2.708(45) & 279.318 & 2.6707(62)~\cite{nicholson2015} & 271.668\\
            & $5s6s~^3\!S_1$ & 1.979(33) & 38.961 & 1.962(10)~\cite{Safronova2013} & 38.26\\
			  & $5s5d~^3\!D_1$ & 2.501(41) & 44.269 & 2.45(24)~\cite{Safronova2013} & 42.45\\
            & $5p~^2~^3\!P_1$ & 2.600(43) & 46.933 & 2.605(26)~\cite{Safronova2013} & 47.096\\
            & $5s7s~^3\!S_1$ & 0.520(9) & 1.717 & 0.516(8)~\cite{Safronova2013} & 1.686\\
            & $5s6d~^3\!D_1$ & 1.172(14) & 7.925 & 1.161(17)~\cite{Safronova2013} & 7.774\\
            & Other &  & 29.72 &  & 29.72\\
            &$\alpha_{\text{core}}$ &  & 5.29 &  & 5.29\\
            & Total &  & 454.1 &  & 443.9\\
		\end{tabular}
	\end{ruledtabular}
    \footnotetext[3]{ In the calculation of polarizabilities, the matrix elements of our theoretical calculation are used.}
    \footnotetext[4]{In the calculation of polarizabilities, the experimental matrix elements are used.}
\end{table*}

Table~\ref{Tab4} presents the static scalar and tensor electric dipole polarizabilities of the $5s^2~^1\!S_{0}$ and $5s5p~^3\!P_{0,1,2}$ states. For the ground state, our calculated results are larger than the other theoretical results. This discrepancy primarily arises because our computed matrix element for the $5s^2~^1\!S_0~\rightarrow~5s5p~^1\!P_1$ transition is larger than the other theoretical values and the experimental values, as shown in Table~\ref{Tab3}. However, when we replace these matrix elements with experimental values, the calculated polarizability agrees very well with the other theoretical results. To illustrate the contribution of the matrix elements, Table~\ref{tab:contribution} shows the breakdown of the contributions from intermediate states to the static polarizabilities. It can be seen that the static polarizability of $5s^2~^1\!S_0$ mainly comes from the contribution of the $5s^2~^1\!S_0~\rightarrow~5s5p~^1\!P_1$ transition. The difference between our calculated transition matrix element of 5.303(74) and the experimental value of 5.248(2) is 1\%. This difference results in a 2\% deviation in the polarizability of $5s^2~^1\!S_0$. Therefore, the error in the polarizability of the ground state $5s^2~^1\!S_0$ is mainly caused by the matrix element of this transition.

For the $5s5p~^3\!P_0$ state, there is an approximately 3\% difference between the our results and the available results. When the recommended values~\cite{Safronova2015}, which were derived from a combined theoretical and experimental analysis of Stark shifts and magic wavelengths, are used, the result agrees very well. In the following calculations of the magic wavelength, these matrix elements were employed. The static tensor polarizability of the $5s5p~^3\!P_{1}$ state, 25.2(0.7) a.u., also shows good agreement with the experimental value of 24.5 a.u.~\cite{oppen1969}. For the $5s5p~^3\!P_{2}$ state, our calculated static tensor polarizability of $-$54.9(1) a.u. agrees well with the relativistic coupled cluster (RCC) result of $-$54.5(2.8) a.u.~\cite{yu2015}, with a difference of 0.7\%. Similarly, for the $5s5p~^3\!P_0$ state, the contribution to the polarizability mainly comes from the $5s5p~^3\!P_0~\rightarrow~5s4d~^3\!D_1$ transition as shown in Table~\ref{tab:contribution}. The difference between our calculated transition matrix element of 2.708(45) and the experimental value of 2.6707(62) is 1.5\%. This difference leads to an approximately 3\% deviation in the polarizability. Therefore, the error in the polarizability of the $5s5p~^3\!P_0$ state is mainly caused by the discrepancy in this transition matrix element.


Table~\ref{Tab5} presents the differential polarizabilities of $\alpha(5s5p~^3\!P_2)-\alpha(5s^2~^1\!S_0)$. These results are compared with those obtained using the RCC method in Ref.~\cite{yu2015}, showing differences of approximately 6\%.
Obviously, this difference is mainly due to the difference in the static polarizability of $^3\!P_2$ states, as shown in Table~\ref{Tab4}. The result of our sum-over-states calculation is 504.3(2.9) a.u., and the result of the finite field method in Ref.~\cite{yu2015} is 491.1(1.0) a.u. However, we can see from Table~\ref{Tab3} that the difference between our calculations of transition matrix elements from the $^3\!P_2$ state and the CI+All method is less than 1\% except for the $5s5p~^3\!P_2~\rightarrow~5s4d~^1\!D_2$ transitions which contributes relatively small to the polarizability. Therefore, the results of the finite field method calculations may underestimate the polarizability of the $^3\!P_2$ state. It is noteworthy that the differential polarizabilities $\Delta\alpha({^3\!P_2~M=\pm2}-{^1\!S_0})$ and $\Delta\alpha({^3\!P_0}-{^1\!S_0})$ are 251.8(2.7) a.u. and 246.3(2.2) a.u., respectively. The difference between these two values is only 5.5 a.u. This slight difference indicates that, at the same temperature, the blackbody radiation (BBR) shifts of the $5s^2~^1\!S_0\rightarrow5s5p~^3\!P_2, M=\pm2$ and $5s^2~^1\!S_0\rightarrow5s5p~^3\!P_0$ transitions are very similar. For example, the BBR shifts for these two transitions at room temperature are $-2.12$ Hz and $-2.16$ Hz, respectively. It should be noted that the BBR shifts of these two clock transitions tend to be almost the same at the same temperature. Therefore, when measurements are performed on the same group of atoms under exactly the same physical conditions (i.e., temperature) and over exactly the same time period, the BBR perturbations experienced by the two transitions may be nearly consistent. In this case, the BBR shifts of the $5s^2~^1\!S_0\rightarrow5s5p~^3\!P_2$ transition can directly refer to the result of the $5s^2~^1\!S_0\rightarrow5s5p~^3\!P_0$ reference clock transition, which may help simplify the error calibration process.

\subsection{Magic wavelengths for linearly polarized light}

 For linearly polarized light, where $\psi=0$ and $\mathcal{A}=0$, $\theta_p=\theta_{maj}$, Eq.~(\ref{eq2}) can be simplified as:
 \begin{eqnarray}\label{eq8}
 	\alpha_i(\omega)=\alpha^S_i(\omega)+\frac{3\text{cos}^2\theta_{maj}}-1{2}\frac{3M^2_{i}-J_i(J_i+1)}{J_i(2J_i-1)}\alpha^T_i(\omega).
 	\nonumber \\
 \end{eqnarray}	
When $J_i \geq 1$, the polarizability includes both scalar and tensor components; however, when $J_i=0$, the polarizability is determined only by the scalar component.

\begin{table}
	\caption[]	{Comparison of the differential polarizabilities $\Delta\alpha({^3\!P_0}-{^1\!S_0})$ and $\Delta\alpha({^3\!P_2}-{^1\!S_0})$. The total polarizability $\alpha(5s5p~{^3\!P_2})$ is calculated by letting $\alpha^V_i=0$ and $\theta_p=0^\circ$ in Eq.~(\ref{eq2}).}\label{Tab5}
	\begin{ruledtabular}
		\begin{tabular}{cll}
			$\Delta\alpha$ & \multicolumn{1}{c}{Present} & \multicolumn{1}{c}{Refs.}  \\
			\midrule
			$\Delta\alpha({^3\!P_0}-{^1\!S_0})$ & 246.3(2.2) & 247.5
                    CI+All~\cite{Safronova2013}\\
                && 244.4(3.5) RCC~\cite{yu2015}\\
			&& 249.8 CI+MBPT~\cite{Porsev2008}\\
			&& 261(4) RMBPT~\cite{porsev2012erratum}\\
			&& 247.379(7) Expt.~\cite{middelmann2012}\\
			$\Delta\alpha({^3\!P_2}-{^1\!S_0})$\\
			$M=0$ & 361.6(3.4)  & 344.2(3.5) RCC~\cite{yu2015} \\
			$M=\pm1$ & 334.2(3.2) & 319.5(3.5) RCC~\cite{yu2015} \\
			$M=\pm2$ & 251.8(2.7) & 236.8(4.1) RCC~\cite{yu2015} \\
		\end{tabular}
	\end{ruledtabular}
\end{table}

\begin{figure}[tbh]	
	\centering{
		\includegraphics[width=8.5 cm, height=17cm]{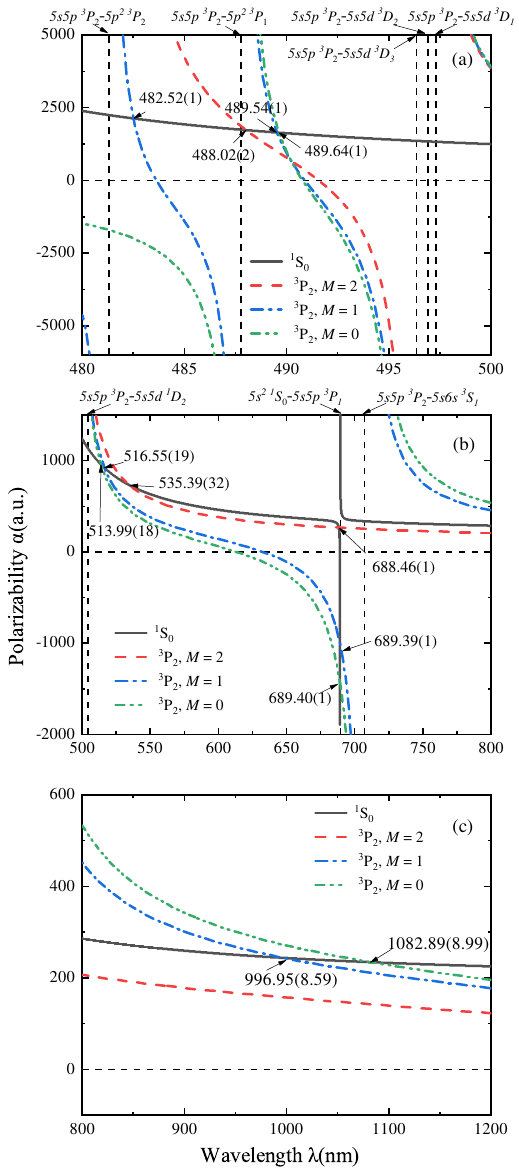} }
	\caption{\label{fig3} Dynamic polarizabilities of the $^1\!S_0$ and $^3\!P_2$ states for linearly polarized light with $\theta_p=0^{\circ}$. The dashed vertical line indicates the positions of the resonant transitions. The magic wavelengths are identified by arrows.}
\end{figure}

\subsubsection{Magic wavelengths for the $5s^2~^1\!S_0$ $\rightarrow$ $5s5p~^3\!P_0$ transition  }

Since the total angular momentum of both the initial and final states in this transition is zero, the polarizabilities each contain only a scalar term. This implies that the magic wavelength is independent of the angles $\theta_p$ and $\theta_k$, as well as the polarization degree $\mathcal{A}$. Figure~\ref{fig2} shows the dynamic polarizabilities of the $5s^2~^1\!S_0$ and $5s5p~^3\!P_0$ states over the wavelength range of $400-1200$~nm. Five magic wavelengths are identified within this range, which are indicated by arrows in the figure. The longest magic wavelength, 813.6(6.2)~nm, agrees well with the experimental value of 813.4280(5)~nm~\cite{ludlow2008}. Most Sr optical lattice clocks use this magic wavelength to trap the Sr atoms. Additionally, the magic wavelength at 497.03(1.09)~nm, located between the resonant transitions $5s5p~^3\!P_0\rightarrow5s5d~^3\!D_1$ and $5s5p~^3\!P_0\rightarrow5s6s~^3\!S_1$, is consistent with the calculated result of 497.0~nm in Ref.~\cite{Safronova2015}. The other three magic wavelengths lie very close to the wavelength of the resonant transition, including 476.92(6)~nm and 432.415(5)~nm magic wavelengths, which are close to the wavelengths of the resonant transitions $5s^2~^1\!S_0\rightarrow5p^2~^3\!P_1$ at 474.324~nm and $5s5p~^3\!P_0\rightarrow5s7s~^3\!S_1$ at 432.766~nm respectively, and the 689.5174(6)~nm magic wavelength is close to the wavelength of the resonant transition $5s^2~^1\!S_0\rightarrow5s5p~^3\!P_1$.

\subsubsection{Magic wavelengths for the $5s^2~^1\!S_0$ $\to$ $5s5p~^3\!P_2$ transition }

For this transition, the total angular momentum of the excited state is two. Its polarizabilities include both scalar and tensor components. The magic wavelength is also angle-dependent. We first calculated the dynamic polarizabilities under the condition where the magnetic field direction is aligned with the laser polarization direction, i.e., $\theta_p=0^{\circ}$. Figure~\ref{fig3} shows the dynamic polarizabilities of the $5s5p~^3\!P_2$ state with $M=0,1,2$. The magic wavelengths are indicated by arrows in this figure and listed in Table~\ref{Tab6}. The corresponding resonant transition positions for both the $5s^2~^1\!S_0$ and $5s5p~^ 3P_2$ states are provided in Table~\ref{Tab6}.
As shown in Fig.~\ref{fig3}(a), when the wavelength is less than 500~nm, the $5s5p~^3\!P_2$ state undergoes numerous resonant transitions that are very close to one another. Consequently, many magic wavelengths occur in the range of $\lambda < 500$~nm. However, these magic wavelengths are close to the resonant transitions, making them unsuitable for magic trapping in experimental settings.

For each magnetic sublevel transition, two sets of magic wavelengths lie between the resonant transitions of the $5s5p~^3\!P_2~\rightarrow~5s5d~^1D_2$ (504.31~nm) and $5s5p~^3\!P_2~\rightarrow~5s6s~^3\!S_1$ (707.20~nm), as illustrated in Fig.~\ref{fig3}(b). The magic wavelengths around 689~nm are very close to the transition wavelength of $5s^2~^1\!S_0 \rightarrow 5s5p~^3\!P_1$ (689.45~nm). Another set of magic wavelengths$-$513.99(18)~nm, 516.55(19)~nm, and 535.39(32)~nm$-$is relatively far from the resonant transitions. Therefore, this set of magic wavelengths can be used for magic trapping of the $5s^2~^1\!S_0$ and $5s5p~^3\!P_2$ states.

Fig.~\ref{fig3}(c) depicts the dynamic polarizability for wavelengths exceeding 800~nm. For the $5s^2~^1\!S_0$ $\to$ $5s5p~^3\!P_2$ with $M=0$ and $M=1$ magnetic sublevel transtions, two magic wavelengths are identified at 1082.89(8.99)~nm and 996.95(8.59)~nm, respectively. These magic wavelengths exhibit sufficient detuning from resonant transitions and lie within the same region as the 813.40-nm magic wavelength of the $5s^2~^1\!S_0~\rightarrow~5s5p~^3\!P_0$ transition. For linearly polarized light at $\theta_p=0^{\circ}$, no magic wavelength is found for the $5s^2~^1\!S_0~\rightarrow~5s5p~^3\!P_2$ $M=2$ clock transition within this range. This absence arises because the polarizabilities of the $5s5p~^3\!P_2$ $M=2$ state remain consistently lower than that of the $5s^2~^1\!S_0$ ground state, as shown in Fig.~\ref{fig3}(c).

It should be noted that the polarizability of the $5s5p~^3\!P_2$ $M=2$ state deviates from that of the $M=0$ and $M=1$ states near the $5s5p~^3\!P_2 \to 5s6p~^3\!S_1$  (707.20~nm) resonant transition, as depicted in Fig.~\ref{fig3}(b). It does not exhibit any singular behavior. This occurs because the contributions from the $5s5p~^3\!P_2 \rightarrow 5s6s~^3\!S_1$ transition to the scalar and tensor components of the dynamic polarizability of the $5s5p~^3\!P_2$ $M=2$ state cancel each other out, resulting in no contribution to the dynamic polarizability of the $5s5p~^3\!P_2$ $M=2$ state. However, when $\theta_p\neq0^{\circ}$, this cancellation is disrupted, giving rise to a singularity in the polarizability of the $5s5p~^3\!P_2$ ($M =2$) state at this resonance. Consequently, for $\theta_p\neq0^{\circ}$, magic wavelengths can be found for the $5s^2~^1\!S_0\rightarrow 5s5p~^3\!P_2$ $M =2$ clock transition within this region.

\begin{table}
\scriptsize
	\caption[]	{Magic wavelengths of the $5s^2~^1\!S_{0}\rightarrow5s5p~^3\!P_{2}$ transition in the case of the magnetic field being vertical to the wave vector $\vec{k}$ with the linearly polarized light, that is $\text{cos}^2\theta_p=1$ and $\mathcal{A} = 0$.}\label{Tab6}
	\begin{ruledtabular}
		\begin{tabular}{cccll}	
			Resonances & $\lambda_{\text{res}}$&  \multicolumn{1}{c}{$|M_i|=0$} &\multicolumn{1}{c}{$|M_i|=1$} &\multicolumn{1}{c}{$|M_i|=2$}\\			
			\midrule
                $5s5p~^3\!P_{2}\rightarrow 5p^2~^3\!P_2$ & 481.32\\
                & & & 482.52(1)  \\
                $5s5p~^3\!P_{2}\rightarrow 5p^2~^3\!P_1$ & 487.77\\
                & & 489.64(1) &  489.54(1) & 488.02(2)\\
                $5s5p~^3\!P_{2}\rightarrow5s5d~^3\!D_3$ & 496.36\\
                & & & 496.90(1) & 496.76(1) \\
                $5s5p~^3\!P_{2}\rightarrow5s5d~^3\!D_2$ &
            496.93\\
                & & 497.29(1)  & 497.30(1) &   \\
                $5s5p~^3\!P_{2}\rightarrow5s5d~^3\!D_1$ &
            497.31\\
                & & & 504.28(1) & 504.21(1)  \\
                $5s5p~^3\!P_{2}\rightarrow5s5d~^1D_2$ &
            504.31\\
                & & 513.99(18)  & 516.55(19) & 535.39(32)  \\
                & & 689.40(1) & 689.39(1) & 688.46(1) \\
                $5s^2~^1\!S_{0}\rightarrow5s5p~^3\!P_1$ & 689.45\\
                $5s5p~{^3\!P}_{2}\rightarrow 5s6s~^3\!S_1$ &707.20\\
			& & 1082.89(8.99) & 996.95(8.59) \\		
		\end{tabular}
	\end{ruledtabular}	
\end{table}

\begin{figure}[tbh]	
 \centering{
 \includegraphics[width=8.5cm, height=7 cm]{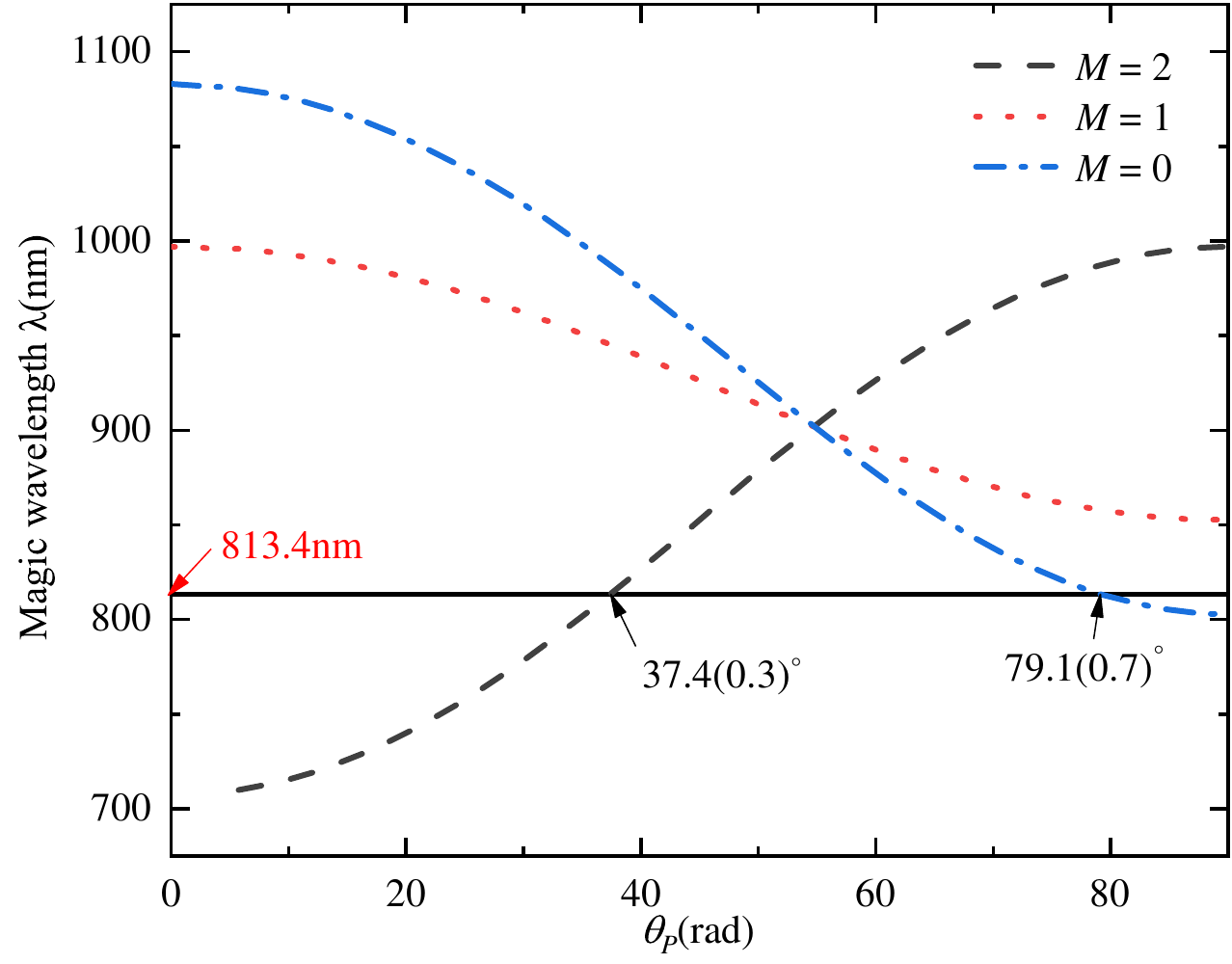} }
 \caption{\label{fig4} The $\theta_p$-dependent magic wavelengths under the linearly polarized light for the $5s^2~^1\!S_0\rightarrow5s5p~^3\!P_2$ transition.The horizontal solid line indicates the position of 813.4~nm.}
 \end{figure}

Figure~\ref{fig4} illustrates the relationship between the longest magic wavelengths and the angle $\theta_p$. Evidently, these magic wavelengths are highly sensitive to $\theta_p$. Notably, for the $5s^2~^1\!S_0~\rightarrow~5s5p~^3\!P_2$ $M=0$ transition, the magic wavelength is 813.4~nm at $\theta_p=79.1(0.7)^\circ$. This indicates that 813.4~nm serves as a common magic wavelength for the $5s^2~^1\!S_0~\rightarrow~5s5p~^3\!P_0$ and $5s^2~^1\!S_0~\rightarrow~5s5p~^3\!P_2$ $M=0$ transitions when the angle $\theta_p$ is set to this specific "magic angle", thereby enabling triple magic trapping. The calculated magic angle shows good agreement with the recent experimental measurement of $78.49(3)^\circ$~\cite{ammenwerth2025realization}, as well as the theoretically predicted value of $75.6^\circ$ in Ref.~\cite{Trautmann2023}. Similarly, 813.4~nm also serves as a common magic wavelength for the $5s^2~^1\!S_0~\rightarrow~5s5p~^3\!P_0$ and $5s^2~^1\!S_0~\rightarrow~5s5p~^3\!P_2$ $M=2$ transitions at $\theta_p=37.4(0.3)^\circ$. In contrast, for the $5s^2~^1\!S_0\rightarrow5s5p~^3\!P_2$ $M =1$ clock transition, no common magic wavelength is found under linearly polarized light. The magic wavelengths corresponding to varying $\theta_p$ are longer than 813.4~nm. Consequently, triple magic trapping for the $5s^2~^1\!S_0$, $5s5p~^3\!P_0$, and $5s5p~^3\!P_2$ $M =1$ states cannot be achieved using linearly polarized light.

\begin{figure*}[tbh]	
\centering{
\includegraphics[width=18cm, height=14cm]{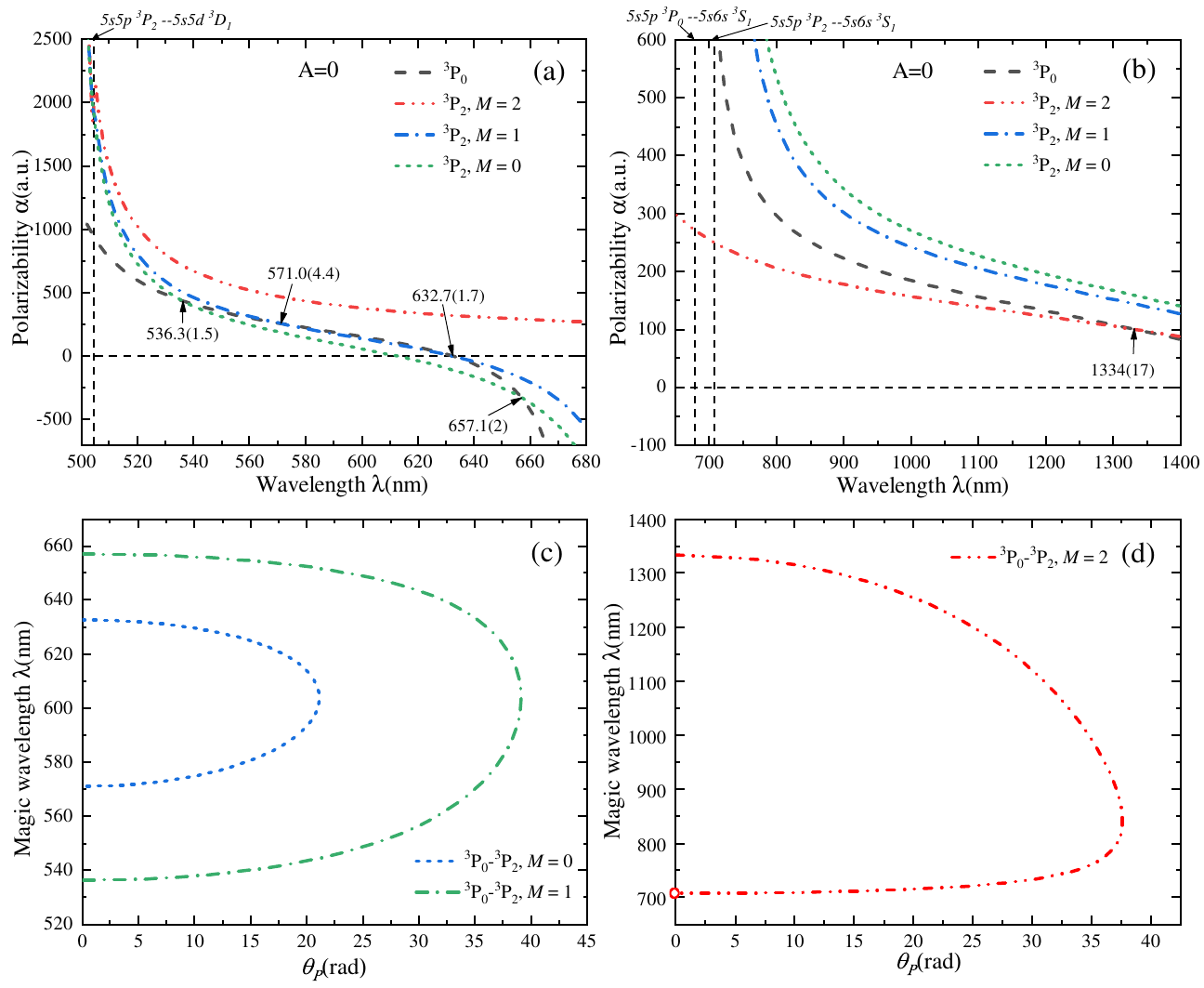} }
\caption{\label{fig3p03p2}  The dynamical polarizabilities under linearly polarized light with $\theta_p = 0^\circ$ and $\theta_p$-dependent magic wavelengths of $5s5p~^3\!P_0$ and $5s5p~^3\!P_2$ states. The vertical dashed lines indicate the positions of the resonant transitions and are given at the top of the figures. }
 \end{figure*}

\subsubsection{Magic wavelengths for $5s5p~^3\!P_0$ $\to$ $5s5p~^3\!P_2$ transition }

The $5s5p~^3\!P_0$ and $5s5p~^3\!P_2$ states of Sr are critically important in quantum computing applications. The realization of magic-wavelength trapping for the $^3\!P_0$ and $^3\!P_2$ states eliminates differential light shift noise caused by fluctuations in the optical lattice laser power. Differential light shifts are a major decoherence mechanism limiting the coherence time of neutral-atom qubits. They also leads to phase drifts and amplitude damping in Rabi oscillations, reducing the fidelity of single-qubit gates. Therefore, the magic-wavelength trapping for the $^3\!P_0$ and $^3\!P_2$ states enhances enhances the qubit coherence and quantum gate fidelity~\cite{pucher2024,unnikrishnan2024}.

Figure~\ref{fig3p03p2} (a) and (b) display the dynamic polarizabilities of the $5s5p~^3\!P_0$ and $5s5p~^3\!P_2$ states under linearly polarized light with $\theta_p = 0^\circ$ for wavelengths $\lambda$>500~nm. Notably, the polarizabilities of $5s5p~^3\!P_0$ and $5s5p~^3\!P_2$ $M=0$, and $M=1$ sublevels are very close to each other. Within the range of 500-700~nm, two magic wavelengths are identified for the transitions $5s5p~^3\!P_0~\rightarrow~5s5p~^3\!P_2~M=0$ and $5s5p~^3\!P_0~\rightarrow~5s5p~^3\!P_2~M=1$. Conversely, no magic wavelengths are found for the $5s5p~^3\!P_0~\to~5s5p~^3\!P_2$ $M=2$ transition within this range. However, one magic wavelength at 1334(17)~nm is found for this transition, as shown in Fig.~\ref{fig3p03p2} (b).

Fig,~\ref{fig3p03p2} (c) and (d) illustrate the dependence of these magic wavelengths on the polarization angle $\theta_p$. For the transition $5s5p~^3\!P_0\rightarrow 5s5p~^3\!P_2$ $M=0$, two magic wavelengths are found for a given $\theta_p$ when $\theta_p<39.1^\circ$, with wavelengths ranging from 536~nm to 657~nm. These magic wavelengths vary continuously as $\theta_p$ changes. In contrast, no magic wavelengths exist for this transition when $\theta_p$>39.1$^\circ$. This absence arises because the polarizability of the $5s5p~^3\!P_2$  $M=0$ state increases with $\theta_p$, causing the polarizabilities of the $5s5p~^3\!P_0$ and $5s5p~^3\!P_2$ $M=0$ states to no longer intersect, resulting in the disappearance of magic wavelengths.
Similarly, for the transition $5s5p~^3\!P_0~\rightarrow~5s5p~^3\!P_2$ $M=1$, two magic wavelengths in the range of 571~nm to 632~nm are present for a given $\theta_p$ when $\theta_p<21.1^\circ$. No magic wavelengths are observed for this transition when $\theta_p>21.1^\circ$. For the transition $5s5p~^3\!P_0~\rightarrow~5s5p~^3\!P_2$  $M=2$, as shown in Fig.~\ref{fig3p03p2} (d), two magic wavelengths within the range of 707~nm to 1334~nm exist for a given $\theta_p$ when 0<$\theta_p<37.56^\circ$. No magic wavelengths are found for this transition when $\theta_p>37.56^\circ$.


 \subsection{The triple magic trapping conditions for the $5s^2~^1\!S_0$, $5s5p~^3\!P_0$ and $5s5p~^3\!P_2$ states }

Since the magic wavelength of the $5s^2~^1\!S_0\rightarrow5s5p~^3\!P_0$ transition is independent of the angles $\theta_p$, $\theta_k$, and the polarization degree $\mathcal{A}$, achieving triple magic trapping for the states $5s^2~^1\!S_0$, $5s5p~^3\!P_0$, and $5s5p~^3\!P_2$ requires selecting a magic wavelength from those associated with the $5s^2~^1\!S_0 \to 5s5p~^3\!P_0$ transition. Among these, the 813.4-nm magic wavelength appears to be the optimal choice, as other candidate magic wavelengths lie too close to the resonant transitions to be viable for trapping.

The triple magic conditions for linearly polarized light are discussed in Section III-B. This section primarily focuses on the conditions for circularly polarized light and arbitrarily elliptically-polarized light.

\subsubsection{Circularly polarized light }

\begin{figure}[tbh]	
\centering{
\includegraphics[width=8.5cm, height=7cm]{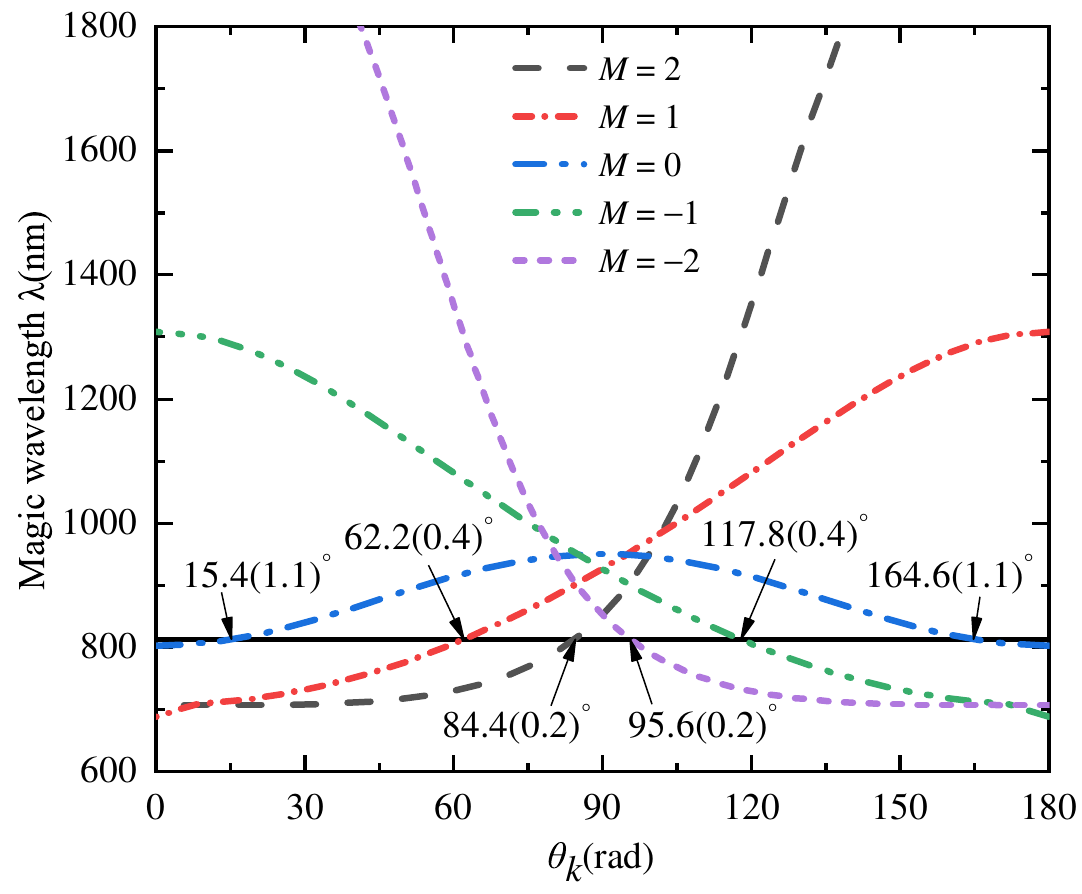} }
\caption{\label{fig5}  $\theta_k$-dependent magic wavelengths under the circularly polarized light for the $5s^2~^1\!S_0\rightarrow5s5p~^3\!P_2$ transition. The horizontal solid line indicates the position of 813.4~nm. The arrow indicates the magic angle that satisfies the triple magic trapping conditions.}
\end{figure}

According to Eq.~(\ref{eq2}), the dynamic polarizabilities of the negative magnetic sublevels under left-handed circularly polarized light are identical to those of the positive magnetic sublevels under right-handed circularly polarized light. Therefore, we will only discuss the results for right-handed circularly polarized light. In this case, with $\psi=\pi/4$, $\mathcal{A}=1$, and $\text{cos}^2\theta_p=\frac{1}{2}\text{sin}^2\theta_{k}$,  Eq.~(\ref{eq2})  simplifies to:
\begin{eqnarray}\label{eq9}
 	\alpha_i(\omega)&=&\alpha^S_i(\omega)+ \frac{\text{cos}\theta_k M_{i}}{2J_i}\alpha^V_i(\omega)  \nonumber \\
 	&+&\frac{\frac{3}{2}\text{sin}^2\theta_k-1}{2}\frac{3M^2_{i}-J_i(J_i+1)}{J_i(2J_i-1)}\alpha^T_i(\omega).
\end{eqnarray}
The polarizability of the $5s5p~^3\!P_2$ state depends on the angle $\theta_k$, making the magic wavelength $\theta_k$-dependent. Figure~\ref{fig5} presents the $\theta_k$-dependent magic wavelengths for the $5s^2~^1\!S_0\rightarrow5s5p~^3\!P_2$ transition under right-handed circularly polarized light. The magic wavelengths corresponding to transitions involving the positive magnetic sublevels are symmetric to those involving the negative magnetic sublevels with respect to the line $\theta_k = 90^\circ$.
When $\theta_k$ = $62.2(0.4)^\circ$ and $84.4(0.2)^\circ$, the magic wavelengths for the $5s^2~^1\!S_0$ $\rightarrow$ $5s5p~^3\!P_2$ transition with the $M=1,2$ sublevels are both 813.4~nm. This indicates that triple magic trapping can be achieved using 813.4~nm circularly polarized light.  For transitions involving the $5s5p~^3\!P_2$ $M=-1,-2$ subleveles, the magic angles are 117.8(0.4)$^\circ$ and 95.6(0.2)$^\circ$, respectively. For the transition involving $5s5p~^3\!P_2$ $M=0$ sublevel, there are two complementary magic angles, namely $15.4(1.1)^\circ$ and 164.6(1.1)$^\circ$.

\subsubsection{Elliptically polarized light}

\begin{figure}[tbh]	
\centering{
\includegraphics[width=8cm, height=16cm]{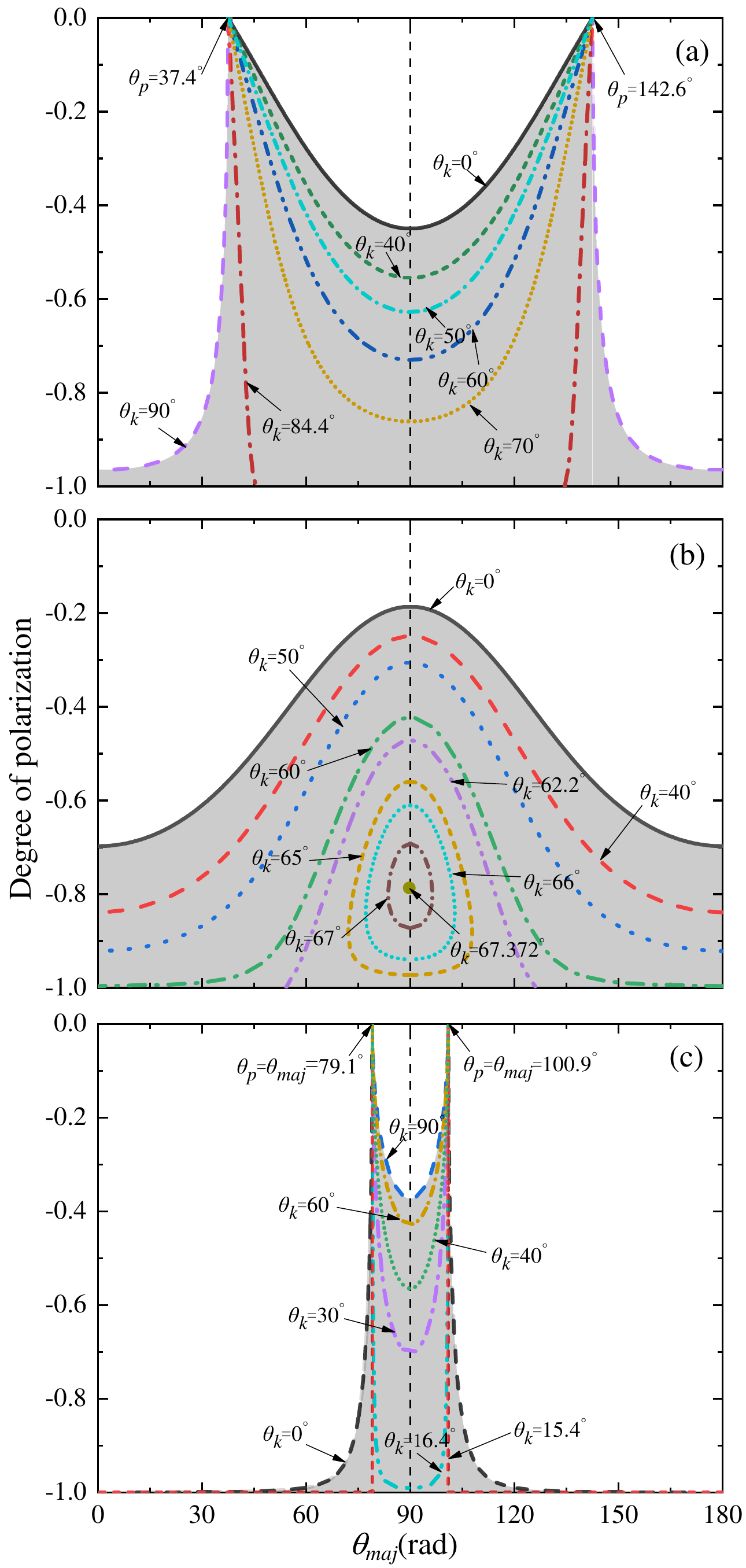} }
\caption{\label{fig6} Triple magic trapping conditions for $5s^2~^1\!S_0\rightarrow 5s5p~^3\!P_{0}$ and $5s^2~^1\!S_0\rightarrow 5s5p~^3\!P_2~M=0, -1, -2$ clock transitions under 813.4~nm elliptically polarized light. The gray area illustrates the parameters $\mathcal{A}$, $\theta_k$, and $\theta_\text{maj}$ satisfying the triple magic trapping conditions (a) illustrates the triple magic trapping conditions for $5s^2~^1\!S_0$, $5s5p~^3\!P_{0}$ and $5s5p~^3\!P_{2}$ $M=-2$ states; (b) depicts that for the $5s^2~^1\!S_0$, $5s5p~^3\!P_{0}$ and $5s5p~^3\!P_{2}$ $M=-1$ states; (c) shows that for the $5s^2~^1\!S_0$, $5s5p~^3\!P_{0}$ and $5s5p~^3\!P_{2}$ $M=0$ states.}
\end{figure}

According to Eqs. (\ref{eq2}) and (\ref{eq3}), the dynamic polarizability for elliptically polarized light can be written as:
\begin{eqnarray}\label{eq10}
   	\alpha_i(\omega)=&&\alpha^S_i(\omega)+ \mathcal{A}\frac{\text{cos}\theta_k M_{i}}{2J_i}\alpha^V_i(\omega)
   	\nonumber \\
   	&&+\frac{3(\frac{1-\sqrt{1-\mathcal{A}^2}}{2}\text{sin}^2\theta_{k}+\sqrt{1-\mathcal{A}^2}\text{cos}^2\theta_\text{maj})-1}{2}
   	\nonumber \\
   	&&\times\frac{3M^2_{i}-J_i(J_i+1)}{J_i(2J_i-1)}\alpha^T_i(\omega).
\end{eqnarray}
For a given magnetic sublevel $M_i$, the polarizability depends on $\mathcal{A}$, $\theta_k$, and $\theta_\text{maj}$.
According to our theoretical calculations, the dynamic scalar, vector, and tensor polarizabilities of the states $5s^2~^1\!S_0$, $5s5p~^3\!P_0$, and $5s5p~^3\!P_2$ at 813.4~nm are $\alpha^S_{^1\!S_0}(\omega)=281.255$ a.u., $\alpha^S_{^3\!P_2}(\omega)=345.368$ a.u., $\alpha^T_{^3\!P_2}(\omega)=-143.787$ a.u., and $\alpha^V_{^3\!P_2}(\omega)=-604.554$a.u., respectively. To satisfy the triple magic trapping conditions $\alpha{(^1\!S_0)} =\alpha{(^3\!P_0)} =\alpha{(^3\!P_2)}$, the following expression can be derived based on the polarizability formula:
\begin{eqnarray}\label{eq11}
	64.113&=&151.139\mathcal{A} M_i\text{cos}\theta_k+143.787\frac{M_i^2-2}{2}
	\nonumber\\
	&\times&\frac{3(\frac{1-\sqrt{1-\mathcal{A} ^2}}{2}\text{sin}^2\theta_{k}+\sqrt{1-\mathcal{A} ^2}\text{cos}^2\theta_\text{maj})-1}{2}. \nonumber\\
\end{eqnarray}
In an experiment, one can adjust the degree of polarization $\mathcal{A}$ or rotate the angles of $\theta_k$ and $\theta_\text{maj}$ to search for triple magic trapping. When these three parameters ($\mathcal{A}$, $\theta_k$, and $\theta_\text{maj}$) satisfy Eq.(\ref{eq11}), triple magic trapping can be achieved.

In Figure~\ref{fig6}, the gray area represents the range where triple magic trapping at 813.4~nm can be achieved for the $5s^2~^1\!S_0$, $5s5p~^3\!P_0$, and $5s5p~^3\!P_2$ states with magnetic sublevels $M = 0$, $-1$, and $-2$. This figure shows only the case where the degree of elliptical polarization $\mathcal{A} <0$. For right-handed elliptically polarized light ($\mathcal{A} > 0$), the range of the triple magic conditions is symmetric to that of $\mathcal{A}~<~0$ with respect to the horizontal axis $\mathcal{A}=0$. However, the value of $\theta_k$ becomes the complementary angle of those in the case of $\mathcal{A} < 0$. When involving the positive magnetic sublevels of the $5s5p~^3\!P_2$, $M=1, 2$ sublevels, the triple magic conditions are similar to those for $M=-1,-2$, except that the polarization direction of the light is reversed.

To facilitate experimental implementation, we suggest setting the angle $\theta_\text{maj}$ to $90^\circ$, i.e., the magnetic field direction lies within the plane formed by the $\hat{\varepsilon}_\text{min}$ and $\vec{k}$ axes. Under this configuration, triple magic trapping can be achieved by adjusting both the angle $\theta_k$ and the light polarization degree $\mathcal{A}$.
For triple magic trapping involving the $5s5p~^3\!P_2$ state with $M=-2$, these two parameters should satisfy:
\begin{eqnarray}\label{eq12}
  272.013&=&-604.554\mathcal{A}\text{cos}\theta_k  \nonumber\\
	&&+215.681(1-\sqrt{1-\mathcal{A}^2})\text{sin}^2\theta_k,
\end{eqnarray}
where the light polarization degree is constrained to $-1\leq\mathcal{A}\leq -0.45$, and $\theta_k$ ranges from $0^\circ\leq\theta_k\leq84.46^\circ$.
For the case involving  the $5s5p~^3\!P_2$ state with $M=-1$, these two parameters should satisfy:
 \begin{eqnarray}\label{eq13}
	 56.333&=&-302.277\mathcal{A}\text{cos}\theta_k   \nonumber\\
     &&-107.84(1-\sqrt{1-\mathcal{A}^2})\text{sin}^2\theta_k,
\end{eqnarray}
with light polarization degree in the range $-1\leq\mathcal{A}\leq -0.186$, and $\theta_k$ restricted to $0^\circ\leq\theta_k\leq67.372^\circ$.
For triple magic trapping involving the $5s5p~^3\!P_2$ state with $M=0$, the parameters satisfy:
\begin{eqnarray}\label{eq14}
        &&15.561=215.681(1-\sqrt{1-\mathcal{A}^2})\text{sin}^2\theta_k,
\end{eqnarray}
where the polarization degree is $-1\leq\mathcal{A}\leq-0.371$, and $\theta_k$ ranges from $15.58^\circ\leq\theta_k\leq90^\circ$.
For this situation, we further recommend experimentally setting $\theta_{maj}=79.1(0.7)^\circ$ and $\theta_k=15.4(1.1)^\circ$. In this case, the triple magic trapping is no longer dependent on the light's polarization degree $\mathcal{A}$, as shown in Fig.~\ref{fig6} (c).

\section{Conclusions}\label{Sec.IV}

The energy levels and electric dipole transition matrix elements of Sr atoms have been calculated using the RCI+MBPT method. The static and dynamic polarizabilities of the  $5s^2~^1\!S_0$ and $5s5p~^3\!P_{0,2}$ states have also been calculated by means of the sum-over-states approach. Our results are consistent with the existing theoretical and experimental results. The magic wavelengths for the transitions of $5s^2~^1\!S_0\rightarrow5s5p~^3\!P_0$, $5s^2~^1\!S_0\rightarrow5s5p~^3\!P_2$, and $5s5p~^3\!P_0\rightarrow5s5p~^3\!P_2$ have been determined. A detailed analysis of how the magic wavelength varies with the angles, $\theta_p$ or $\theta_k$, is also provided.

In addition, the triple magic trapping conditions at 813.4~nm for the $5s^2~^1\!S_0$, $5s5p~^3\!P_0$, and $5s5p~^3\!P_2$ states under different polarized lights have also been calculated. For linearly polarized light, triple magic trapping can be achieved at $\theta_p =79.1(0.7)^\circ$ involving the $5s5p~^3\!P_2$ $M=0$ state, and at $\theta_p =37.4(0.3)^\circ$ involving the $5s5p~^3\!P_2$ $M=\pm2$ states. The conditions for achieving triple magic trapping with elliptical polarization involving different magnetic sublevels are also provided. We suggest that in the experiment, the angle $\theta_\text{maj}$ be set to $90^\circ$, i.e., the direction of the magnetic field lies within the plane formed by the $\hat{\varepsilon}_\text{min}$ and $\vec{k}$ axes. In this case, triple magic trapping can be achieved by adjusting both the angle $\theta_k$ and the light polarization degree $\mathcal{A}$. These two parameters should be within the ranges $-1\leq\mathcal{A}\leq -0.45$ and $0^\circ\leq\theta_k\leq84.46^\circ$ involving the $^3\!P_2$ $M=-2$ state; $-1\leq\mathcal{A}\leq -0.186$ and  $0^\circ\leq\theta_k\leq67.372^\circ$ involving $^3\!P_2$ $M=-1$ state; and $-1\leq\mathcal{A}\leq -0.371$, and  $15.58^\circ\leq\theta_k\leq90^\circ$ involving $^3\!P_2$ $M=0$ state. When involving the $^3\!P_2$ $M=0$ state, we further recommend setting the angles $\theta_{maj}$ and $\theta_k$ to $79.1(0.7)^\circ$ and $15.4(1.1)^\circ$, respectively.  In this case, the triple magic trapping condition is independent of the degree of light polarisation  $\mathcal{A}$.

\section{Acknowledgments}
This work has been supported by the National Key Research and Development Program of China (Grant No. 2022YFA1602500), the National Natural Science Foundation of China (Grant Nos. 12174316, 12174268, 12364034, 12464036, and 12404306), the Natural Science Foundation of Gansu Province (No. 25JRRA799), and the key Talent Project of Gansu Province (No.2025RCXM021)

%

\end{document}